\documentclass[aps,prx,twocolumn,footinbib, superscriptaddress, nofootinbib]{revtex4-2}

\usepackage{graphicx}
\usepackage{hyperref}
\usepackage{amsmath,amssymb, mathtools, amsthm}
\usepackage{bbold}
\usepackage{tikz} 
\usetikzlibrary{quantikz}
\usepackage{orcidlink}
\usepackage{microtype}
\usepackage[T1]{fontenc}

\begin{document}
	
	\title{Ensemble-Based Quantum Token Protocol Benchmarked on IBM Quantum Processors}
	
	\author{Lucas Tsunaki \orcidlink{0009-0003-3534-6300}}
	\affiliation{Department Spins in Energy Conversion and Quantum Information Science (ASPIN), Helmholtz-Zentrum Berlin für Materialien und Energie GmbH, Hahn-Meitner-Platz 1, 14109 Berlin, Germany}	
	
	\author{Bernd Bauerhenne \orcidlink{0000-0002-3397-2290}}
	\affiliation{Institute of Physics and Center for Interdisciplinary Nanostructure Science and Technology (CINSaT), University of Kassel, Heinrich-Plett-Strasse 40, 34132 Kassel, Germany}
	
	\author{Malwin Xibraku \orcidlink{0009-0002-3183-6188}}
	\affiliation{Institute of Physics and Center for Interdisciplinary Nanostructure Science and Technology (CINSaT), University of Kassel, Heinrich-Plett-Strasse 40, 34132 Kassel, Germany}
	
	\author{Martin E. Garcia \orcidlink{0000-0003-2418-1902}}
	\affiliation{Institute of Physics and Center for Interdisciplinary Nanostructure Science and Technology (CINSaT), University of Kassel, Heinrich-Plett-Strasse 40, 34132 Kassel, Germany}
	
	\author{Kilian Singer \orcidlink{0000-0001-9726-0367}}
	\email{ks@uni-kassel.de}
	\affiliation{Institute of Physics and Center for Interdisciplinary Nanostructure Science and Technology (CINSaT), University of Kassel, Heinrich-Plett-Strasse 40, 34132 Kassel, Germany}
	
	\author{Boris Naydenov \orcidlink{0000-0002-5215-3880}}
	\email{boris.naydenov@helmholtz-berlin.de}
	\affiliation{Department Spins in Energy Conversion and Quantum Information Science (ASPIN), Helmholtz-Zentrum Berlin für Materialien und Energie GmbH, Hahn-Meitner-Platz 1, 14109 Berlin, Germany}
	\affiliation{Berlin Joint EPR Laboratory, Fachbereich Physik, Freie Universität Berlin, 14195 Berlin, Germany}
	
	\date{\today}
	
	\begin{abstract}
		Quantum tokens envision to store unclonable quantum states in a physical device, with the goal of being used for personal authentication protocols, as required by banks.
		Still, the experimental realization of such devices faces many technical challenges, which can be partially mitigated using ensembles instead of single qubits.
		In this work, we thus propose an ensemble-based quantum token protocol, describing it through a simple yet general model based on a quantum mechanical observable.
		The protocol is benchmarked on five IBM quantum processors and a general hacker attack scenario is analyzed, in which the attacker attempts to read the bank token and forge a fake one, based on the information gained from this measurement.
		We experimentally demonstrate that the probability that the bank erroneously accepts a forged coin composed of multiple tokens can reach values below $10^{-22}$, while the probability that the bank accepts its own coin is above 0.999.
		The overall security of the protocol is therefore demonstrated within a hardware-agnostic framework, confirming the practical viability of the protocol in arbitrary quantum systems and thus paving the way for future applications with different ensembles of qubits, such as color center defects in solids.
	\end{abstract}
	
	\maketitle
	
	
	\section{Introduction}\label{sec:intro}
	
\begin{figure*}[t!]
		\includegraphics[width=\textwidth]{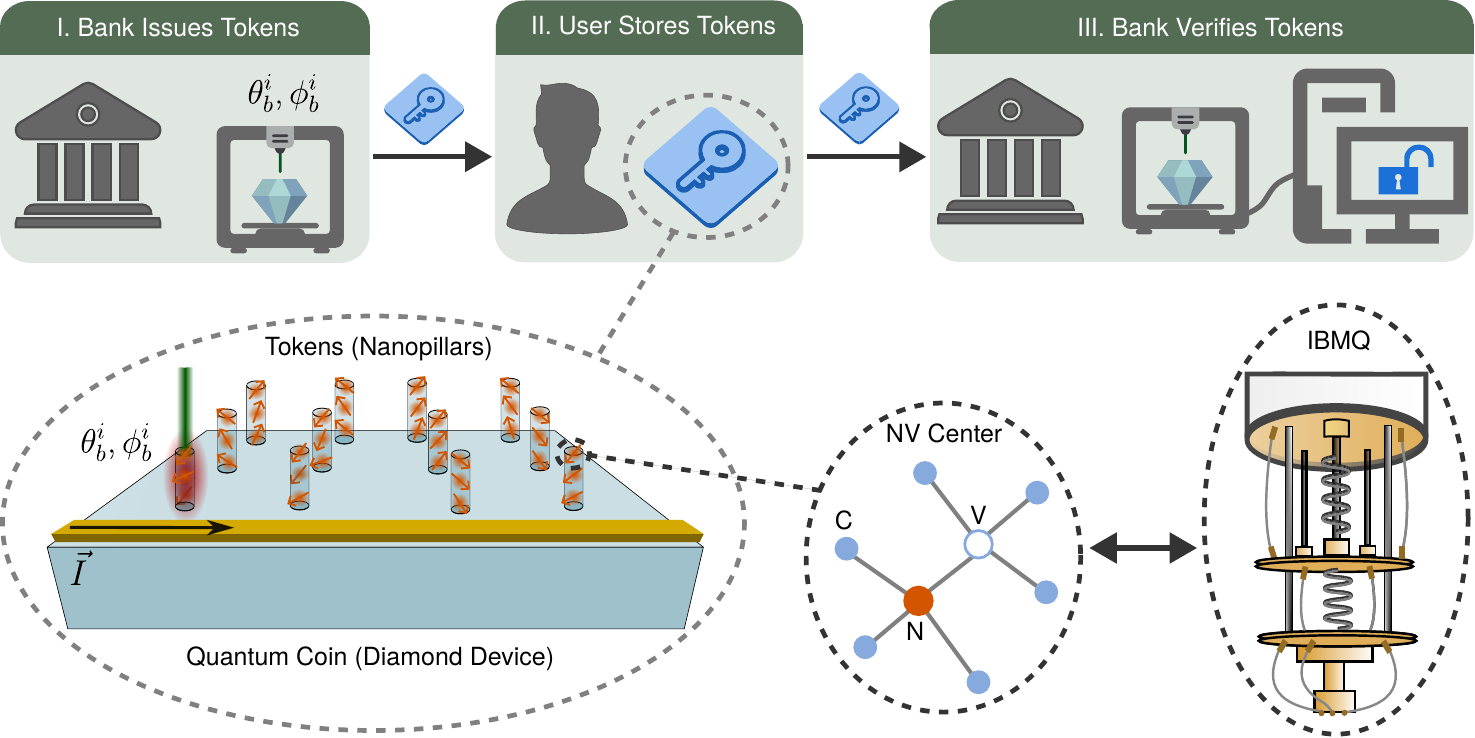}
		\caption{Proposed ensemble-based quantum token protocol and the diamond quantum coin device.
			An issuing agent (e.g. a bank) generates many quantum tokens with angles $\theta_b^i$ and $\phi_b^i$ in a quantum coin device.
			The device is then physically handed to the user, who stores it until authentication by the bank is required, which consists in measuring the tokens in the coin and comparing them with the previously generated angles.
			The coin is accepted if a minimum number of tokens in the coin above a certain security threshold is measured in the right state.
			The physical implementation of the tokens is envisioned to be achieved with diamond nanopillars (for optimal photon collection efficiency~\cite{nanopillars1, nanopillars2}) containing nitrogen-vacancy center ensembles. 
			Whereby adopting tokens composed of ensembles instead of single qubits, the application can be technology less demanding.
			In a combination of optical methods~\cite{optical_pumping} and resonant electromagnetic pulses~\cite{rabi}, the electronic spins of the NVs could be used as auxiliary qubits for measuring~\cite{single_shot_readout} and controlling~\cite{SWAP_NV} the nitrogen nuclear spins, used as memory qubits.
			Nonetheless, such physical implementation of a diamond quantum coin faces many technical challenges regarding material fabrication~\cite{token_fabrication}.
			With that, a benchmark of the protocol within a hardware-agnostic framework is vital for future quantum token applications, which is accomplished with IBM quantum platforms.
		}
		\label{fig:DIQTOK_proposal}
	\end{figure*}

	Quantum no-cloning theorem~\cite{no_cloning} is the basis for secure exchange of information in quantum communications and cryptography~\cite{quantum_cryptography}.
	Another interesting application is the creation of mobile non-copyable storage units with a natural expiration date, i.e. a quantum token~\cite{quantum_token, patent_quantum_token}.
	The proposed quantum token protocol is schematically presented in Fig.~\ref{fig:DIQTOK_proposal}.
	First, an issuing agent (e.g. a bank) generates many tokens, each in a different quantum state with angles $\theta_{b}^i$ and $\phi_{b}^i$ on the Bloch sphere.
	All quantum tokens compose together a physical device, which we denominate as a quantum coin.
	The device is then physically handed to the user, who stores it until authentication is required.
	To perform the authentication, the bank measures all tokens in the coin and compares them with the angles as they originally prepared in each token.
	If a minimum number of tokens in the coin are measured in the correct state above a predetermined security threshold, the coin is successfully authenticated.
	Otherwise, a cloning attempt is heralded.
	In such a way, the quantum information is stored, transported and later authenticated rather than simply transmitted, as in quantum communication applications~\cite{NV_network, NV_teleportation_1}.
	If successful, such a technology would represent a robust system for the highest security requirements, such as bank cards or personal IDs.
	
	Promising candidates for quantum systems with long coherence, room temperature operation, low power consumption, potential for miniaturization, ease of state manipulation and readout are color centers in transparent solid state substrates~\cite{color_centers}, such as the nitrogen-vacancy (NV) center in diamond~\cite{NV}. When considering these systems, the use of ensembles as quantum tokens rather than single qubits could make this application technologically less demanding~\cite{ensemble_control}. Whereby entering a redundant quantum parallelism regime with many qubits performing the same protocol, the partial decoherence of one or more qubits of the ensemble will not result in a complete failure of the quantum token protocol. However, the conventional single qubit quantum token protocol~\cite{quantum_token} does not allow to use ensembles. As the quantum no-cloning theorem is not trivially applicable for ensembles, which already consist of identical copies of the individual qubits. In this work, we then demonstrate that an ensemble-based quantum token protocol is still safe and reproducible. 
	
	Additional challenges appear when dealing with ensembles rather than single qubits, which can be mitigated or corrected with the appropriate technique. For instance, many optimal quantum control methods, such as composite pulses~\cite{composite_pulses1, composite_pulses2}, are being developed to allow for optimal ensemble manipulation, taking inhomogeneous broadening due to slightly different qubit energies and control field amplitude into account. Another important factor is the ensemble separability, that is, the ability of an attacker to manipulate and measure individual qubits of the ensemble. In this case, by making measurements on different axes, an attacker can more efficiently read and create forged tokens, as discussed in detail in our companion paper~\cite{OURPRA}. For NVs and other color centers in contrast, the ensemble is typically already not separable, due to the diffraction limited optical polarization and readout~\cite{optical_pumping} and to non-local microwave excitation~\cite{rabi} techniques common to them. Nonetheless, our method is still applicable if measurements on sub-ensembles are performed.
	
	\begin{figure*}[t!]
		\centering
		\includegraphics[width=\textwidth]{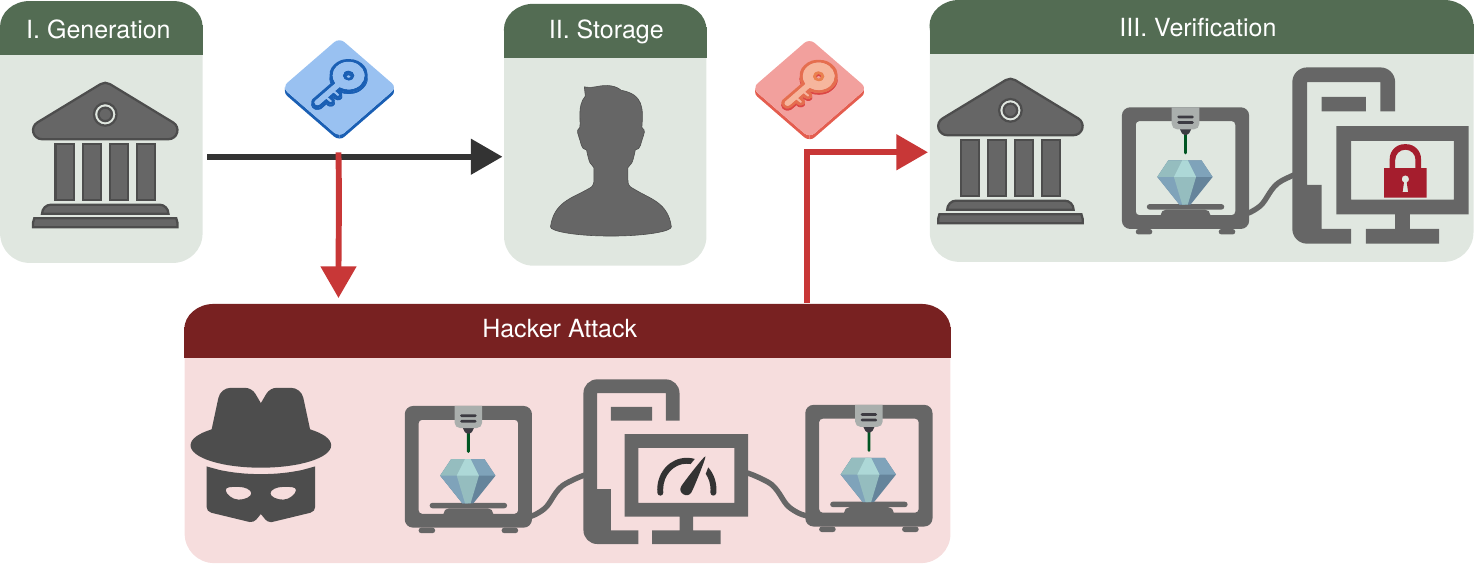}
		\caption{General hacker attack scenario considered in this work.
			Firstly, the attacker attempts to gain information about the quantum coin prepared by the bank by measuring its tokens.
			With this information, they forge fake tokens, which they use to attempt to impersonate the user.
			The bank needs to be able to discern the fake tokens from the ones which they originally prepared, where the disparity between the efficiency in which the bank generates its own tokens and the efficiency in which the attacker forges fake ones defines the security of the quantum token protocol.
		}
		\label{fig:diagram_attack}
	\end{figure*}	
	
	A physical candidate for a quantum coin device can be imagined as series of diamond nanopillars containing small ensembles of NVs (see Fig.~\ref{fig:DIQTOK_proposal}).
	Each nanopillar then represents a quantum token, providing improved photon collection efficiency as compared to bulk diamond~\cite{nanopillars1, nanopillars2}.
	The states of the tokens are optically initialized~\cite{optical_pumping} by the bank and prepared in the electronic spin of the NVs into the states $\theta^i_b$ and $\phi^i_b$ with resonant microwave pulses~\cite{rabi}, transmitted through an antenna micro-fabricated on the diamond surface~\cite{antenna}. The states are then transferred to nitrogen nuclear spins of the NVs or to nearby $^{13}$C with longer coherence times through a SWAP gate~\cite{SWAP_NV}, which could potentially be implemented by dynamical decoupling of the electron spin~\cite{ambiguous_resonances}. The latter can be extended during storage, by also applying dynamical decoupling sequences to the nuclear spins for error mitigation~\cite{nuclear_spin_dd}. In addition, the NV charge can be changed during storage, leading to a spinless electronic state which further extends the nuclear coherence~\cite{charge_state}. Finally, in order to verify the angles $\theta_b^i$ and $\phi_b^i$ of each token in the coin, the bank performs an optical single-shot readout of the nuclear spins using the electron spins as an ancilla qubit~\cite{single_shot_readout}.
	
	The experimental realization of such a diamond-based quantum coin is still an ongoing research.
	More specifically, it is currently technologically limited by fabrication techniques of such a device~\cite{token_fabrication} and the aforementioned control and single shot readout methods for NV ensembles.
	Therefore, it is crucial to benchmark the ensemble token protocol on a hardware-agnostic high-level representation quantum processor, such as the IBM Quantum Platforms (IBMQ).
	This approach allows us to not only identify the quality parameters of the quantum token hardware, but also to quantify the efficiency of the proposed protocol, demonstrating its potential hardware interoperability.
	And lastly, by comparing the efficiency of the bank in preparing and authenticating its own tokens with the efficiency in which a hacker can clone the quantum token, we obtain a measure of the protocol's safety.
	
	In this study, we thus consider one general possible attack scenario, as represented in Fig.~\ref{fig:diagram_attack}. First, the hacker attempts to gain information about the quantum coin prepared by the bank by performing projective measurements to the tokens.
	Using this information, the attacker forges a fake token and passes it to the bank, seeking to impersonate the user.
	During authentication, the bank then needs to be able to discern the forged coin from the one it originally created.
	Although this scenario covers a general framework for many quantum token hardware candidates, there could potentially exist other more efficient attack methods~\cite{side_channel_attack_1, interaction_free_meas}, not covered in this study.
	
	For benchmarking the proposed protocol and quantifying its security against such hacker attacks, we used five different IBMQ superconducting processors~\cite{qiskit, hardware1, hardware2, hardware3}.
	Due to the increased memory costs in waveform generation for reproducing an ensemble of qubits with IBMQ, we invoke the ergodic principle, in which the ensemble average is substituted by a time-average of the same qubit prepared and measured multiple times.
	More experimental details are given in Appendix~\ref{sec:A2}, while the proofs of mathematical equations are all presented in Appendix~\ref{sec:A1}.
	Due to the increased number of mathematical variables in this work, Tab.~\ref{tab:variables} provides a glossary of their definitions.
	All codes for experimental control and result analysis are open-source and provided at~\cite{github}, which also contains a graphical user interface (GUI) where the results presented in this work can be calculated and visualized for the parameters of an arbitrary quantum token hardware.
	
	This work is structured as follows.
	In Sec.~\ref{sec:uncertainty}, we start by presenting a model for the measurement uncertainty which a general quantum token hardware is subject to, being then benchmarked in the IBMQ.
	The highlight of this approach is that it enables us to determine the main quality parameters that will describe most of the tokens' behavior.
	Building on this model, in Sec.~\ref{sec:bank}, the bank generation and authentication protocol is defined and benchmarked, which defines how well can the bank prepare and read its own tokens.
	Finally, in Sec.~\ref{sec:attacker}, we execute the attack method and conduct a detailed analysis of the resulting protocol's security.
	
	
	\begin{figure*}[t!]
		\centering
		\begin{quantikz}[font=\large]	
			\lstick{\ket{0}} & \gate{\hat{R}(\theta_b, 0)} & \meter{$\hat{N}$}
		\end{quantikz}
		\vspace{-1cm}
		
		\includegraphics[width=\textwidth]{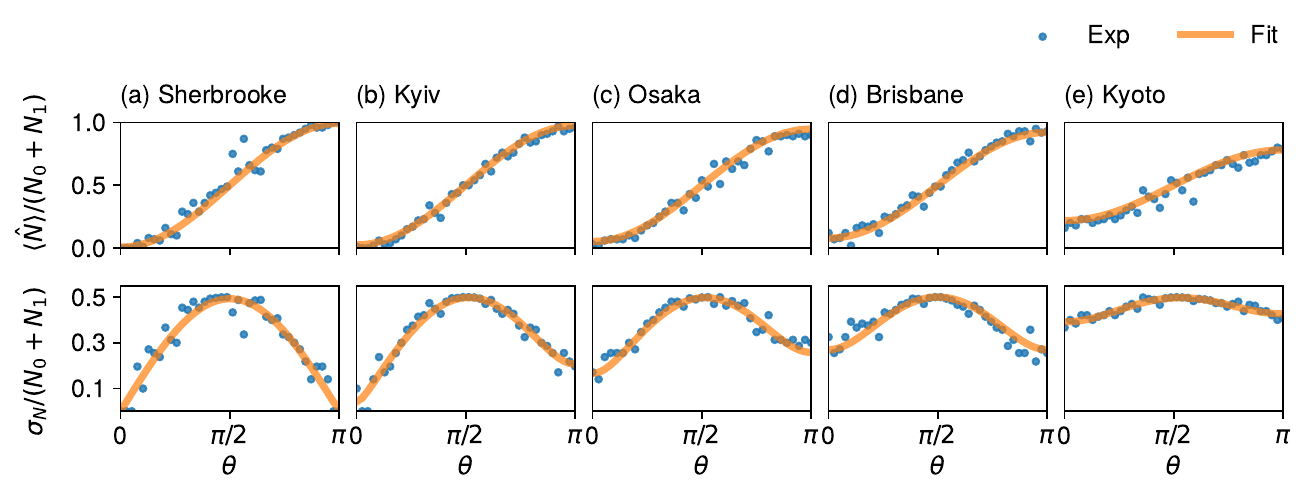}
		\caption{Normalized expectation values $\langle \hat{N} \rangle/(N_0 + N_1)$ and uncertainty $\sigma_N/(N_0 + N_1)$ as a function of $\theta$ measured with IBM Quantum Platforms \textbf{(a)} Sherbrooke, \textbf{(b)} Kyiv, \textbf{(c)} Osaka, \textbf{(d)} Brisbane and \textbf{(e)} Kyoto. $\langle \hat{N} \rangle$ values follow a Rabi oscillation from the state $\ket{0}$ to $\ket{1}$, while the uncertainties are maximum at $\theta = \pi/2$ due to quantum projection noise. Kyiv and Osaka show a strong influence from shot noise, given that the uncertainty is higher at $\theta=\pi$ than at $\theta=0$. On the other hand, Brisbane and Kyoto are dominated by experimental noise, resulting in a constant offset to the the uncertainty. Last, Sherbrooke presents the highest contrast, with almost no contribution from shot and experimental noise. The fitted values of normalized contrast $c$ and $\sigma_{exp}$ are given in Tab.~\ref{tab:fit_params}. }
		\label{fig:noise_analysis}
	\end{figure*}
	
	\section{Uncertainty Analysis}\label{sec:uncertainty}
	
	A primary obstacle in modeling a quantum processor as the IBMQ is the lack of complete access to its specifications, due to proprietary constraints. A crucial parameter that can be leveraged to extract valuable information about a physical system is its inherent uncertainty or noise, characterized by its statistical properties~\cite{FDT}. In addition to that, a noise based framework is common to other quantum system, respecting the protocols' universality. As will be shown, this uncertainty is never zero even without any experimental errors, due to quantum uncertainty principle
	
	For the description of the ensemble system we use a simple Hamiltonian-independent model that describes a general quantum state on the Bloch-sphere by the two angles $\theta$ and $\phi$, additionally to expressions for expectation values and variance of the ensemble measurements.
	Rigorously speaking, the state of the ensemble is in fact in a statistical mixture, described by a density matrix instead of a pure ket state.
	However, to simplify the theoretical treatment, we assume pure states for the quantum token ensemble - an approximation which holds notably well with the IBMQ, as will become clear in the following sections.
	In most physical systems, the angles $\theta$ and $\phi$ are not directly measured. Instead, the bank has an observable $\hat{N}$ given by
	\begin{equation*}
		\hat{N} = \begin{bmatrix}
			N_0 & 0 \\
			0 & N_1 \\
		\end{bmatrix},
	\end{equation*} 
	which can be used to measure the populations in the $\ket{0}$ and $\ket{1}$ states, where $N_0, N_1 \in \mathbb{R}$. This observable could be related to some elaborate quantity like the change of magnetic flux or charge in a superconducting circuit, as with IBMQ. But for a more concrete physical example, we take this observable related to a photon count, as is the case with NVs. For a general Bloch state with angles $\theta$ and $\phi$, the expectation value of $\hat{N}$ is
	\begin{equation}\label{eq:expectation}
		\langle \hat{N} \rangle = N_0 \cos^{2}{\left(\frac{\theta}{2} \right)} + N_1 \sin^{2}{\left(\frac{\theta}{2} \right)} ,
	\end{equation}
	not depending on $\phi$. Ideally, the bank would have one of the eigenvalues as 0. In reality however, if we assume $\ket{1}$ to be the bright state such that $N_1>N_0$, $N_0$ is related to the background counts and $N_1$ to the collection efficiency of the measurement. Opposite to the IBMQ, in NVs we have $N_1<N_0$, which does not affect the end results of the quantum token. As it will become clear in Sec.~\ref{sec:bank} and~\ref{sec:attacker}, the quality of a quantum token can be characterized in terms of the normalized contrast between $N_1$ and $N_0$, $c\equiv(N_1-N_0)/(N_0+N_1)$. The universality of the protocol extends to systems which support a measurement scheme that allows for the measurement of an ensemble average $\hat{N}$, a set of universal rotations $\hat{R}(\theta, \phi)$ and can be approximately described by pure states.
	
	A general assumption can be made that the total uncertainty $\sigma_N$ for a measurement of $\hat{N}$ is composed of three factors~\cite{error_2}. Firstly, any quantum system is subject to a quantum projection noise, or Heisenberg uncertainty, given by the operator variance $\sigma_{q}^2 = \langle \hat{N}^2 \rangle - \langle \hat{N} \rangle^2$. This value is highest when the states are in a maximum superposition with $\theta=\pi/2$ and zero when they are in the eigenbasis states, with $\theta=0$ or $\theta=\pi$. Secondly, given the quantized nature of the emitted photons, following a Poisson statistics, the total uncertainty also has a contribution from shot noise as $\sigma_{s}^2=\langle \hat{N} \rangle$. Contrarily to the quantum projection noise, the shot noise is maximum at $\theta=\pi$. Finally, we also consider a generic Gaussian experimental error given by a constant $\sigma_{exp}$, adding an offset to the uncertainty. This error can be defined to incorporate pulses error in the state preparation and other experimental noise sources. Assuming that these three components are statistically independent, they can be added in quadrature~\cite{error_1}, resulting in
	\begin{align}\label{eq:uncertainty}
		\sigma_N^2 & = \sigma_q^2 + \sigma_n^2 + \sigma_{exp}^2 \nonumber \\
		& = N_0 \cos^{2}{\left(\frac{\theta}{2} \right)} (1 + N_0) + N_1 \sin^{2}{\left(\frac{\theta}{2} \right)} (1 + N_1) \nonumber \\
		& - \left[N_0 \cos^{2}{\left(\frac{\theta}{2} \right)} + N_1 \sin^{2}{\left(\frac{\theta}{2} \right)}\right]^{2} + \sigma_{exp}^2.
	\end{align}
	Depending on the values of $N_0$, $N_1$ and $\sigma_{exp}$, the total noise will be dominated by either one of the three components. Solutions of $\sigma_N$ for different combinations of values are shown in Fig.~\ref{fig:sigma_analytical} in Appendix~\ref{sec:A1}. We also provide a GUI application~\cite{github}, which can be used to visualize $\sigma_N$ for arbitrary parameters input by the user.
	
	To test this noise model, a Rabi measurement~\cite{rabi_original, rabi} for $\langle \hat{N} \rangle$ was performed on the IBMQ as a function of $\theta$ for $\phi=0$, while $\sigma_N$ is taken from the standard deviation of 100 shots. To better compare the data between the different quantum processors, both values are normalized by $N_0+N_1$. The high-level representation of the quantum circuit of the experiment and the results are shown in Fig.~\ref{fig:noise_analysis}. By fitting $\langle \hat{N} \rangle$ with Eq.~\ref{eq:expectation} and $\sigma_N$ with Eq.~\ref{eq:uncertainty}, we obtain the values of the normalized contrast $c$ and $\sigma_{exp}$ (Tab.~\ref{tab:fit_params}).
	
	All five quantum processing units show a maximum of $\sigma_N$ at $\theta=\pi/2$, resulting from a strong contribution from quantum projection noise to the total uncertainty. In Kyiv and Kyoto, a large contribution from shot noise is also observed, given by a higher uncertainty at $\theta=\pi$ over $\theta=0$. While in Brisbane and Kyoto the experimental error is the dominant term, adding a constant offset to the uncertainty. Finally, Sherbrooke has the highest contrast and is completely dominated by quantum projection noise, with negligible contribution from shot noise and experimental error. Which can be explained in terms of the hardware's slightly lower median readout and gate errors, apart from longer coherence times.	
	
	This prominent quantum projection noise can be used as an extra security measure for the bank. As an increased noise would indicate that the token is being measured in the wrong axis, which in turn would herald a cloning attempt by an attacker. However, in cases with inefficient read out - as typically is the case with NV color centers - shot noise can dominate quantum projection noise. In those cases, our protocol is still applicable, but more quantum tokens per coin are needed~\cite{OURPRA}.
	
	\begin{table}[t!]
		\begin{tabular}{|@{\hspace{.1cm}}c@{\hspace{.1cm}}||@{\hspace{.2cm}}c@{\hspace{.2cm}}|c|@{\hspace{.2cm}}c@{\hspace{.2cm}}|@{\hspace{.2cm}}c@{\hspace{.2cm}}|@{\hspace{.2cm}}c@{\hspace{.2cm}}| }
			\hline
			IBMQ & $c$ & $\sigma_{exp}/_{N_0+N_1}$ & $\bar{n}_b$ & $\bar{n}_f$ & $p_f$ \\ \hline
			Sherbrooke & 0.986  & $10^{-5}$ & 0.990 & 0.685 & 0.065 \\ 
			Kyiv    & 0.950  & 0.026 & 0.992 & 0.682 & 0.059 \\
			Osaka    & 0.896  & 0.158 & 0.970 & - & - \\ 
			Brisbane & 0.843  & 0.270 & 0.879 & 0.611 & 0.285 \\ 
			Kyoto    & 0.563  & 0.377 & 0.792 & - & - \\ \hline
		\end{tabular}
		\caption{Quality parameters of the token hardware. The values of normalized contrast $c$ and experimental uncertainty $\sigma_{exp}$ are obtained from the fit of $\langle \hat{N} \rangle$ and $\sigma_N$ (Sec.~\ref{sec:uncertainty}), with the first representing how easy is to determine the state of the system and the second is defined to incorporate all additional experimental errors. The mean experimental bank self-acceptance $\bar{n}_b$ represents the fraction of qubits in a token which the bank prepares and measures in the correct state (Sec.~\ref{sec:bank}). Analogously, $\bar{n}_f$ is the fraction of qubits which the bank accepts from a forged token (Sec.~\ref{sec:attacker}). Finally, $p_f$ is the probability of acceptance of one forged token if we set the acceptance threshold $n_T$ such that the bank acceptance probability is $p_b>0.999$. Overall, Sherbrooke and Kyiv have the best quality parameters, such that a minor improvement of $c$ in comparison to Brisbane leads to a significant increase in the token security. Forgery measurements could not be concluded using Osaka and Kyoto due to the systems' retirement in August 2024.}
		\label{tab:fit_params}
	\end{table}
	
	
	\section{The Bank Protocol}\label{sec:bank}
	
\begin{figure}[t!]
		
		\begin{quantikz}[font=\large]
			\lstick{\ket{0}} & \gate{\hat{R}(\theta_b, \phi_b)} & \gate{\hat{R}^{-1}(\theta_b, \phi_b)} & \meter{$\hat{N}$} 
		\end{quantikz}
		
		\includegraphics[width=\columnwidth]{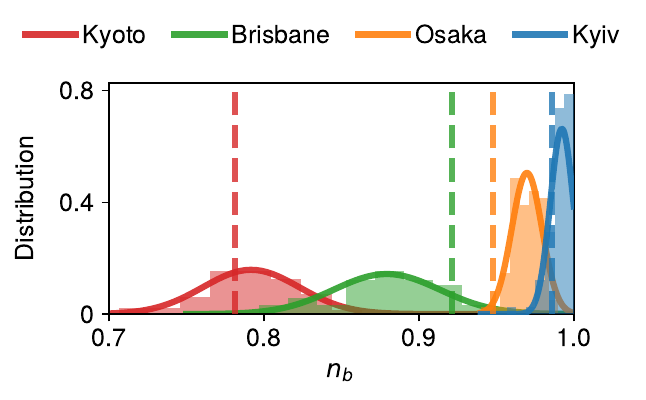}
		\caption{Self acceptance fraction $n_b$ distributions of IBM Quantum Platforms Kyiv, Osaka, Brisbane and Kyoto measured for several combinations of linearly distributed angles $\theta_{b}$ and $\phi_b $. $n_b$ is defined as the number of qubits in a token which are measured in the state $\ket{0}$ after a bank rotation of $\hat{R}(\theta_b, \phi_b)$, followed by an inverse rotation and measurement of $\hat{N}$. The distributions are fitted with Gaussian functions, determining the acceptance range of each hardware, with a mean $\bar{n}_b$ roughly around $N_1/(N_0 +N_1)$ (dashed lines). As the contrast of the hardware increases, the mean values of $n_b$ get higher and the linewidth of the distributions get narrower. Sherbrooke results are omitted, as they overlap with Kyiv results.}
		\label{fig:nb_benchmark}
	\end{figure}

	The next important quantity to be benchmarked is the fraction of qubits which the bank prepares and reads successfully, without interference of an attacker. We assume a protocol as depicted in Fig.~\ref{fig:nb_benchmark}. Starting from the $\ket{0}$ state, the bank prepares a state with a rotation $\hat{R}(\theta_b, \phi_b)$. Subsequently, for authentication, the bank unrotates the token with the inverse operation $\hat{R}^{-1}(\theta_b, \phi_b)$ and makes a measurement on $\hat{N}$, yielding\footnote{Or $n_b=\langle \hat{N} \rangle/(N_0 + N_1)$, if we had chosen $\ket{1}$ as the initial reference state or if $N_0 > N_1$.} $n_b=1-\langle \hat{N} \rangle/(N_0 + N_1)$ qubits of the token at the initial  state $\ket{0}$, where both the counts and the uncertainty are smallest.
	
	This acceptance fraction is highly dependent on the specific pulse errors of the quantum hardware. Still, some general conclusions can be drawn just from the
	eigenvalues of the observable $\hat{N}$. If no rotation errors were made and there was no decoherence, the two rotations would cancel and the final state would be $\ket{\Psi_f} = \hat{R}^{-1}(\theta_b, \phi_b) \hat{R}(\theta_b, \phi_b) \ket{0} = \ket{0}$. Although this is not the experimental reality, it gives a rough estimate on the average self-acceptance fraction of
	\begin{equation}\label{eq:max_nq}
		\bar{n}_b \approx  1 - \frac{\bra{0} \hat{N} \ket{0}}{N_0 + N_1} = \frac{N_1}{N_0 +N_1} .
	\end{equation}
	Clearly, the higher the contrast between the dark counts $N_0$ and the bright counts $N_1$, the easier it is for the bank to distinguish the state. This also sets a limit for the self-acceptance fraction.
	Therefore, the bank should choose a threshold for the photon count $n_T$ such that only tokens which show $n_b > n_T$ are accepted in the coin. 
	
	The acceptance fraction $n_b$ was benchmarked in the five IBMQ for different combination of linearly distributed angles $\phi_b$ and $\theta_b$. In Fig.~\ref{fig:nb_benchmark}, the $n_b$ distributions of each hardware are shown and fitted with Gaussian functions, defining the acceptance range of each token. For clearer visualization, the results from Sherbrooke are omitted in the figure, as they completely overlap with Kyiv, but can be found online~\cite{github}. The distributions present good agreement with their values of $N_1/(N_0 +N_1)$ fitted from $\langle \hat{N} \rangle$ and $\sigma_N$ (Fig.~\ref{eq:uncertainty} and Tab.~\ref{tab:fit_params}). As the normalized contrast $c=(N_1-N_0)/(N_0+N_1)$ of the hardware increases, not only the mean values $\bar{n}_b$ get higher, but the linewidth of the distributions get smaller. With Kyiv and Sherbrooke showing the highest self-acceptance fractions, while Kyoto has the worst, in accordance to the results from Sec.~\ref{sec:uncertainty}. The $n_b$ values as a function of $\theta_{b}$ and $\phi_b $ are shown in Fig.~\ref{fig:nb_angles}, where no angle dependence is observed. In hardware scenarios with non-optimized long pulses and short coherence times, we would expect smaller acceptance fractions for larger angles due to increased errors in longer pulses. Which is not the case in these IBMQ, given the high fidelity of their qubit operations.
	
	
	\section{An Attack Scenario}\label{sec:attacker}
	
	The last aspect which defines the quality of a quantum token hardware is how much better the bank can prepare and accepts its own tokens compared to tokens forged by an attacker. We imagine one general attack scenario where the attacker makes a measurement of the bank token, then uses the obtained information to forge a fake one and pass it to the bank. The attack scenario is schematically represented in Fig.~\ref{fig:DIQTOK_proposal}.
	
\begin{figure*}[t!]
		\centering
		\begin{quantikz}[font=\large]
			\lstick{\ket{0}} & \gate{\hat{R}(\theta_b, \phi_b)} & \gate{\hat{R}^{-1}(\theta_{a}, 0)} & \meter{$\hat{N}$}
		\end{quantikz}
		
		\includegraphics[width=\textwidth]{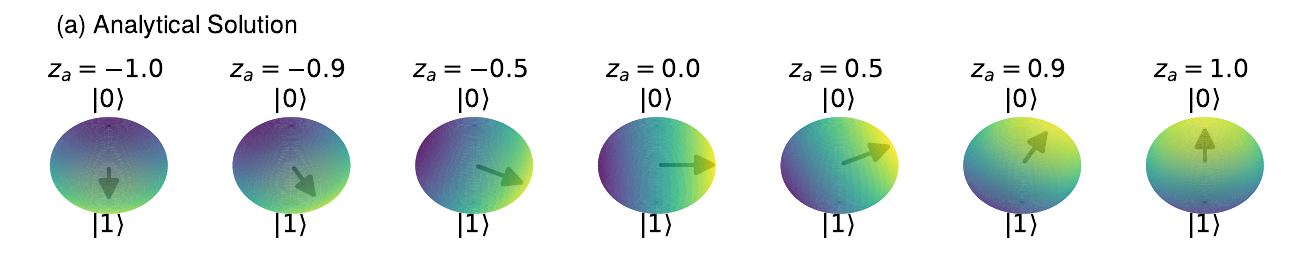}
		\includegraphics[width=\textwidth]{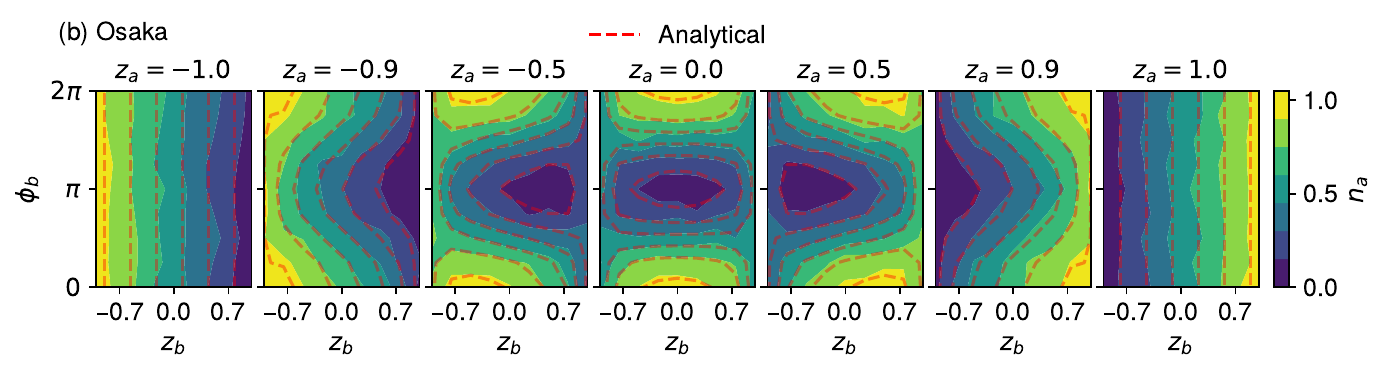}
		
		\caption{\textbf{(a)} Analytical solution and \textbf{(b)} experimental data from IBM Quantum Platform Osaka for the fraction of qubits in a token which are measured in the state $\ket{0}$ by an attacker $n_a$ as a function of the bank preparation angles $z_b = \cos \theta_{b}$, $\phi_b$ and the attacker measurement angle $z_a = \cos \theta_a$  for $\phi_a=0$. The analytical solution for (a) is taken from Eq.~\ref{eq:na_sol} with $c=1$, while the red-dashed lines in (b) show the solution for $c$ from Tab.~\ref{tab:fit_params}. The strong agreement of the analytical model with the experimental data evidences that the hardware can be well described only in terms of the operator $\hat{N}$ and its uncertainty. Clearly, the closer the measurement angles are from the bank angles, the higher $n_a$ will be. By making a measurement attempt, the attacker is able to conclude that the bank is in a line of the Bloch sphere solution of the parametric equation resulting from Eq.~\ref{eq:na_sol} with fixed $n_a$. The attacker then uses this knowledge to forge a fake token and pass it to the bank.}
		\label{fig:na_benchmark}
	\end{figure*}

	As discussed in the previous section, the bank prepares the token with a rotation $\hat{R}(\theta_b, \phi_b)$. In succession, the attacker unrotates with some chosen angles $\theta_a$ and $\phi_a$, leading to a final state as $\ket{\Psi_f} = \hat{R}^{-1}(\theta_a, \phi_a )\hat{R}(\theta_b, \phi_b) \ket{0}$. Lastly, the attacker measures the observable $\hat{N}_a$, which in practice is not necessarily the same observable the bank would use $\hat{N}_a \neq \hat{N}$, as they do not have the same experimental setup to measure the token. However, for the sake of simplicity we consider the case where the faker can make their measurements as accurate as the bank, with $\hat{N}_a = \hat{N}$, which is the case in the IBMQ. Thus, analogously to Eq.~\ref{eq:max_nq}, the fraction of qubits $n_a$ on the token which are measured in the state $\ket{0}$ by the attacker is 
	\begin{equation}\label{eq:na_sol}
		2 n_a = 1 + c [ \cos{\theta_a}\cos{\theta_b} + \sin{\theta_a}\sin{\theta_b}\cos{(\phi_b - \phi_a)} ] .
	\end{equation}
	Integrating and averaging over the bank angles $\theta_{b}$ and $\phi_b$ (Appendix~\ref{sec:A1}), we obtain $\bar{n}_a = 1/2$, which does not depend on $c$. In this way, $n_a$ is centered around 1/2 and $c$ corresponds to a normalized contrast, as expected. The analytical solution of $n_a$ for $c=1$ is plotted over the Bloch sphere in Fig.~\ref{fig:na_benchmark}~(a), where the arrows represent the measurement vectors of the attacker with $z_a=\cos \theta_a$. As observed, the closer the attacker measurement angles $\theta_{a}$ and $\phi_a$ are from the bank token $\theta_{b}$ and $\phi_b$, the highest is the value of $n_a$ measured by the attacker. The two pole measurements at $z_a=-1$ and +1 are not symmetric, due to the increased shot noise at $\ket{1}$. The GUI application~\cite{github} also provides visualization of $n_a$ solution for arbitrary user defined parameters.
	
	This attacker measurement scheme was performed on the five IBMQs as a function of the angles $z_b = \cos(\theta_b)$, $\phi_b$ and $z_a = \cos(\theta_a)$ for a fixed $\phi_a$. The resulting angle dependence of $n_a$ is shown for IBMQ Osaka with $\phi_a=0$ in Fig.~\ref{fig:na_benchmark}~(b), with the Bloch sphere being projected into a plane for better visualization. The results for the other four quantum processors are shown in Fig.~\ref{fig:na_SM}, with measurement angles $\phi_a=0$ and $\phi_a=\pi/2$. The data is also compared with the analytical expression from Eq.~\ref{eq:na_sol} for the value of $c$ fitted from the uncertainty measurement (Tab.~\ref{tab:fit_params}). In all cases, the measurements show a strong agreement with the analytical formula, further indicating that the quality of a quantum token can be fully characterized in terms of the $\hat{N}$ operator eigenvalues.
	
	\begin{figure*}[t!]
		\begin{quantikz}[font=\large]
			\lstick{\ket{0}} & \gate{\hat{R}(\theta_b, \phi_b)} & \gate{\hat{R}^{-1}(\theta_{a}, \phi_a)} & \meter{$\hat{N}$}  & \xRightarrow{n_a}  &  &\lstick{\ket{0}} & \gate{\hat{R}(\theta_f, \phi_f)} & \gate{\hat{R}^{-1}(\theta_b,\phi_b)} & \meter{$\hat{N}$}
		\end{quantikz}
		
		\centering
		\includegraphics[width=\textwidth]{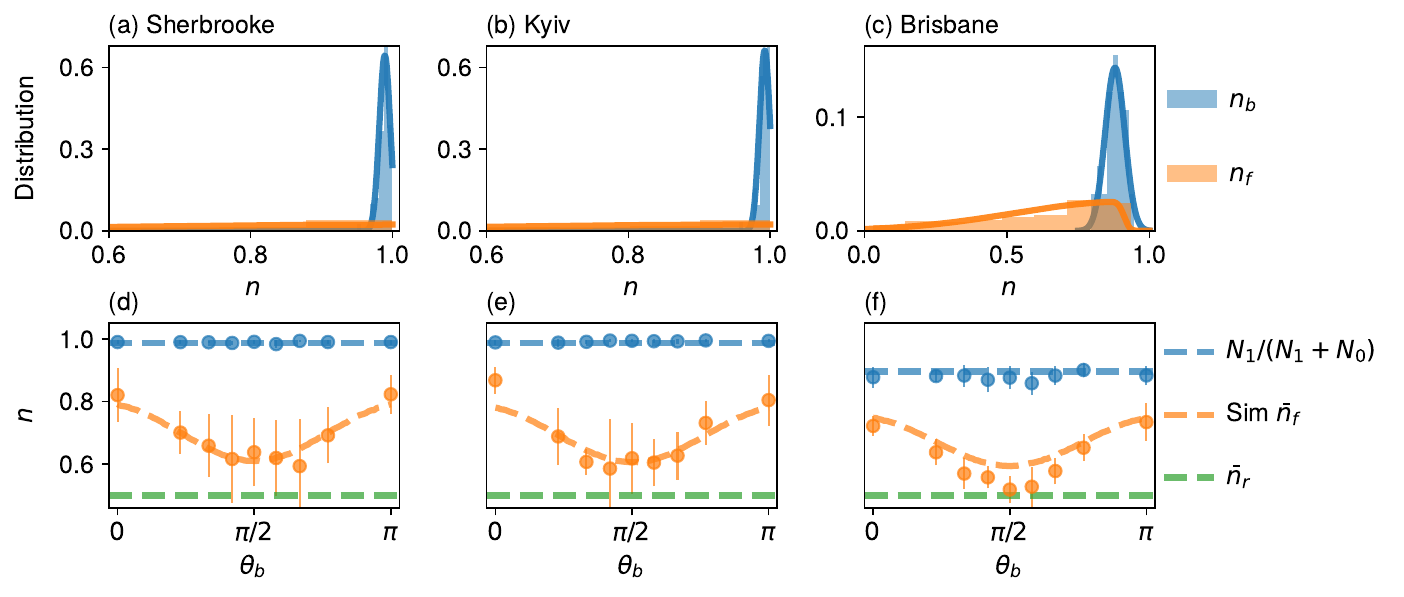}
		\caption{Forged tokens distributions $n_f$ compared to the bank distributions $n_b$ measured in IBM Quantum Platforms \textbf{(a)} Sherbrooke, \textbf{(b)} Kyiv and \textbf{(c)} Brisbane. First, the attacker makes a measurement as described in Fig.~\ref{fig:na_benchmark}. Using the measured values of $n_a$, they try to forge a fake token and pass to the bank with angles $\theta_f$ and $\phi_f$, which the bank will measure resulting in $n_f$ qubits in the $\ket{0}$ state. The $n_f$ distributions follow a phenomenological skew normal distribution, with a large standard deviation and negative scale. Most of its values are above the threshold for randomly generate tokens $\bar{n}_r =1/2$ (green dashed line), indicating that the forger procedure is robust, but still worse than the verification procedure of the bank. In Sherbrooke and Kyiv, the forged tokens distributions are completely overshadowed by the narrow bank distribution. \textbf{(d)}, \textbf{(e)} and \textbf{(f)} The angle dependence of $n_f$ with $\theta_{b}$ indicates that the bank tokens are more easily cloneable at the poles rather than the equator, due to a smaller state distribution there. The $n_f$ distributions are compared with their simulated values [Fig.~\ref{fig:acceptance} (a)], while the $n_b$ distributions are compared with their expected values of $N_1/(N_1 + N_0)$.}
		\label{fig:nf_brisbane}
	\end{figure*}
	
	Based on the measured value of $n_a$, the attacker can deduce the possible angles of the bank token $\theta_b$ and $\phi_b$ by solving Eq.~\ref{eq:na_sol}. This results in a solution line over the Bloch sphere, as visualized in Fig.~\ref{fig:na_benchmark}. However, due to the ambiguity of the solution, the attacker cannot unequivocally determine which point in the line with just one measurement. In this study, we consider a case where the attacker uses the information obtained from the measurement and randomly chooses one of the solutions to generate a forged token with angles $\theta_f$ and $\phi_f$, which is already a much more robust attack than randomly guessing a point in the whole Bloch sphere. More efficient methods where the attacker can subdivide the ensemble into parts and measure each of them in different axes are discussed in detail in our companion paper~\cite{OURPRA}, even though this is not a realistic scenario with most color center systems (Sec.~\ref{sec:intro}).
	
	If the attacker chooses the measurement angle as $\theta_{a}=0$ or $\pi$, then $\sin \theta_a = 0$ and Eq.~\ref{eq:na_sol} can be simply inverted leading to a solution for the angle of the forged token as
	\begin{equation*}
		\theta_f = \arccos \left( \frac{2 n_a -1}{\cos \theta_{a} c}\right).
	\end{equation*}
	In this case, $\theta_f$ is completely determined, while $\phi_f$ is undetermined and the attacker must choose an arbitrary value. This can be observed for $z_a=\pm1$ measurements shown in Fig.~\ref{fig:na_benchmark}~(b), where $n_a$ is completely independent of $\phi_b$. It may be the case, however, that $\theta_f$ has no real solution due to experimental errors and thus the attacker simply guesses the angles. If instead, the attacker does not measure the token at the poles, one can solve Eq.~\ref{eq:na_sol} for $\phi_b$, which gives two solutions
	\begin{equation*}
		\phi_f = \phi_a \pm \arccos \left[ \frac{(2n_a-1)/c - \cos \theta_a \cos \theta_f}{\sin \theta_a \sin \theta_f} \right] .
	\end{equation*}	
	In order to obtain a valid solution for $\phi_f$, the argument of the inverse cosine function must be located in the interval $[-1,1]$. This provides a condition for valid $\theta_f$ values as shown in Eq.~\ref{eq:zf_interval}. After randomly choosing a $\theta_f$ value inside the interval, the attacker takes one of the two solutions of $\phi_f$. This procedure for forging a fake token is illustrated in Fig.~\ref{fig:zf_interval}.
	
	The proposed token forgery protocol was implemented for different combinations of attack angles ($\theta_a$, $\phi_a$) and linearly distributed bank angles ($\theta_b$, $\phi_b$), based on the $n_a$ values measured in Fig.~\ref{fig:na_SM}. This results in a fraction of $n_f$ qubits in the token which are measured in the $\ket{0}$ state by the bank, where $n_f$ is symmetrical to Eq.~\ref{eq:na_sol} substituting the attacker measurement angles by the forged angles $a \rightarrow f$. In this way, if an attacker randomly generates fake tokens, they will get $\bar{n}_r =1/2$ fraction of qubits accepted by the bank, independent of $c$. The experimental distributions of $n_f$ are shown in Fig.~\ref{fig:nf_brisbane} for (a) Sherbrooke, (b) Kyiv and (c) Brisbane\footnote{The measurements on Osaka and Kyoto could not be completed due to their retirement in August 2024.}. Unlike the bank self-acceptance fraction $n_b$, the forger distributions do not follow a Gaussian function. Instead, we phenomenologically model them by a skew normal distribution~\cite{skew_pdf} with a large standard deviation and negative shape. Is evident that the forger is able to more efficiently generate forged tokens than the average for randomly guessing $\bar{n}_r =1/2$, thus demonstrating the robustness of the forger protocol. Still, the distribution from the forged tokens are below the bank self-acceptance distributions. Where in Sherbrooke and Kyiv, the behavior of $n_f$ is completely dominated by the narrow $n_b$ distributions.
	
	By examining the values of $n_f$ as a function of the bank polar angle $\theta_b$ [Figs.~\ref{fig:nf_brisbane}~(d), (e) and (f)], we observe that the attacker has a larger chance of success if the bank tokens are closer to the poles rather than at the equator of the Bloch sphere. This is a direct result from the probability distribution of the angle $\theta$ in spherical coordinates system $f_\theta(\theta)=\sin (\theta)/2$, being larger at $\theta = \pi/2$. Meaning that there are more distribution of states in the equator than at the poles for the attacker to guess. Thus, the bank should follow such a distribution and prepare more tokens at the equator, differently from these measurements. The experimental values of $n_f$ are also in good agreement with simulation, considering the contrast values $c$ obtained from the uncertainties fit (Sec.~\ref{sec:uncertainty}). Small discrepancies between the simulated and experimental values can be attributed to contrast parameter estimation errors.
	
\begin{figure*}[t!]
		\includegraphics[width=\textwidth]{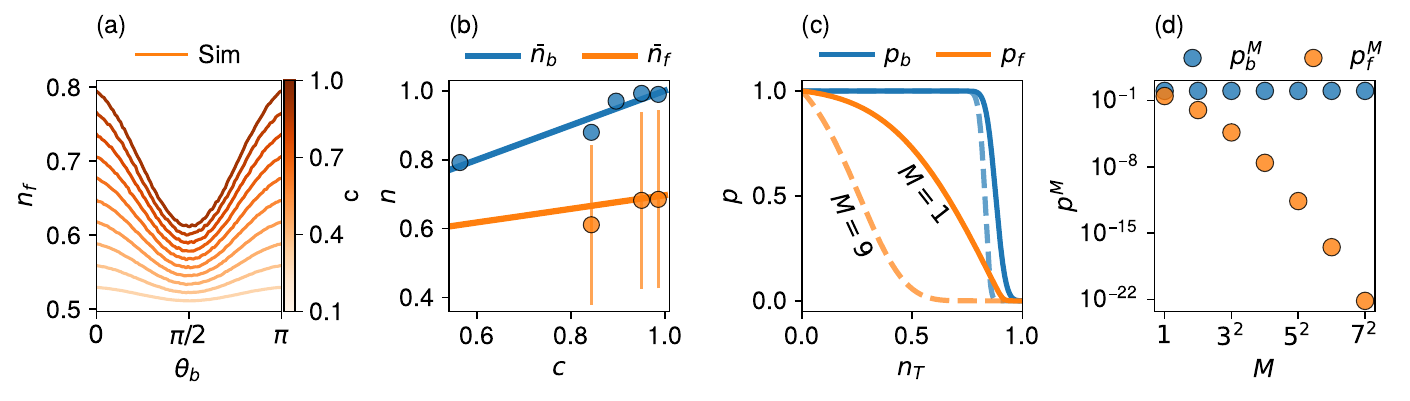}
		\caption{(\textbf{a)} Simulated values of $n_f$ for different normalized contrasts $c$. The mean values of $n_f$ increase with $c$, indicating that the token is more easily cloneable. \textbf{(b)} Simulated forger acceptance fractions $\bar{n}_f$ and bank self-acceptance $\bar{n}_b = N_1/(N_1+N_0)$ as a function of the normalized contrast $c$. Experimental results for the five IBM Quantum Platforms are also shown, with good agreement with the model. Error bars for $n_b$ values are smaller than the points, while $n_f$ present large standard deviations due to the nature of the forger protocol. As $c$ increases, both values increase linearly, but faster for the bank. Meaning that the protocol becomes safer. \textbf{(c)} Acceptance probability for the forger $p_f$ and the bank $p_b$ calculated from the fitted distributions as a function of the acceptance threshold $n_T$ for 1 and 9 tokens in Brisbane. With a single token, $p_f$ is non-negligible for $n_T$ around 0.8, but with the increase in the number of tokens the chance of acceptance of a fake token approaches zero. \textbf{(d)} Acceptance probabilities $p_f^M$ and $p_b^M$ of IBMQ Brisbane as a function of the number of tokens in the coin $M$. The acceptance threshold is chosen such that $p_b^M>0.999$. This shows that $p_f^M$ can be made negligible with just a few tokens.}
		\label{fig:acceptance}
	\end{figure*}

	Additional simulations for different contrast values $c$ are shown in Fig.~\ref{fig:acceptance}~(a), indicating an increase of $n_f$ with $c$. In Fig.~\ref{fig:acceptance}~(b), the simulated mean values of $\bar{n}_f $ are compared with the experimental data from IBMQ, showing a linear dependence with $c$. The $n_f$ distributions have intrinsically large standard deviations, due to the nature of the forger protocol. In addition, the experimental values of $n_b$ (Sec.~\ref{sec:bank}) are compared with its theoretical values from Eq.~\ref{eq:max_nq}. The results show that the tokens become more easily clonable as the contrast of the hardware increases, given that the attacker can better rely on their measurement of the bank token. On the other hand, $\bar{n}_b \approx N_1 /(N_1 +N_0)$ is also linearly dependent on $c$, but with a larger linear coefficient. Which in the end, makes the token safer.
	
	The danger of the bank token being successfully cloned is related to the intersection of the $n_b$ and $n_f$ distributions in Fig.~\ref{fig:nf_brisbane}~(a), (b) and (c). Therefore, the bank should carefully choose the acceptance threshold $n_T$ such that most of its own tokens are accepted, while forged tokens are rejected. We define the probability of acceptance $p_b$ of the bank tokens and $p_f$ of the forged tokens as the integral of the fitted distributions above the acceptance threshold $n_T$. These acceptance probabilities can be visualized and calculated for arbitrary user defined parameters in our GUI application as well~\cite{github}.
	
	The bank has the freedom to adjust the acceptance threshold $n_T$ such that its own acceptance probability $p_b$ is set to desired values, while the forger acceptance probability is minimum. Fig.~\ref{fig:acceptance}~(c) shows both acceptance probabilities as function of $n_T$ in IBMQ Brisbane, the hardware with the lowest contrast. If the bank chooses a value of $p_b>0.999$, this results in $p_f$ values for the three IBMQ which are shown in Tab.~\ref{tab:fit_params}. This demonstrates that a small improvement of less than 15\% in contrast from Brisbane to Kyiv results in a reduction of $p_f$ by a factor of almost 5. This significant increase in security with just a small improvement in the token quality is caused by the highly non-linear behavior of the acceptance probability, being proportional to the overlap area between the two $n_f$ and $n_b$ distributions. Therefore, as the difference between the two distributions increase linearly with $c$ [Fig.~\ref{fig:acceptance}~(b)] and the linewidth of $n_b$ gets smaller, the resulting token security rapidly increases.
	
	Another important quantity for the security of the protocol is the number of tokens in the coin $M$. Likewise, it should also be optimized by the bank to increase security, as the acceptance probability of the entire coin becomes $p^{M}$. In Fig.~\ref{fig:acceptance}~(c), we observe that the acceptance probability of a forged token is non-negligible for $M=1$. However, if the number of tokens in the coin is increased to $M=9$, $p_f^9$ is greatly reduced, while the probability of acceptance for a bank $p_b^9$ coin is less affected. To further demonstrate this, Fig.~\ref{fig:acceptance}~(d) shows the acceptance probability $p^{M}_f$ of IBMQ Brisbane for different values of $M$, where $n_T$ is chosen such that the acceptance probability of the bank tokens are always above $p^M_b > 0.999$. $M$ values are taken imagining a square lattice of tokens in the coin device. This shows that with a reduced number of tokens smaller than $M=7^2=49$, is already possible to obtain acceptance probabilities of forged tokens below $10^{-22}$, while keeping high acceptances for the bank verification. Therefore, simply by increasing the number of tokens in a coin, the security of the protocol can be enhanced to desired levels, even with hardware of poor performance.

	
	\section{Conclusion and Outlook}
	
	Overall, this study proposes and realizes an ensemble-based quantum token protocol by benchmarking it on the IBM superconducting processors. The experimental noise characterization of the quantum hardware permits to calculate the eigenvalues and uncertainty of the observable $\hat{N}$. Where the large demonstrated quantum projection noise can be used as an extra protection layer, heralding a cloning effort if the token is measured in the wrong axis. In turn, the contrast between the dark and bright states $c$ almost entirely defines the quality of the quantum coin, as observed in the bank self-acceptance and forger measurements. The model's independence from the exact Hamiltonian of the system and its other intrinsic properties demonstrates the hardware-agnosticism of the protocol. Furthermore, the high agreement between theory and the experiments show that the model is complete, but still general to a large class of quantum systems that support a measurement observable operator $\hat{N}$ and a set of universal rotations $\hat{R}(\theta, \phi)$.
	
	Our protocol also shows great potential with the ongoing quantum hardware evolution, as a minor improvement in the hardware quality represented by the contrast $c$ leads to significant improvements in security. But even for hardware with low performance, it was shown that the security can be increased to arbitrarily high standards simply by increasing the number of tokens in the coin, within realistic values. Additional security can obtained by preparing more states in the equator than at the poles of the Bloch sphere, due to the increased proficiency for an attacker to forge tokens prepared at the poles. We further provide an open-source tool with graphical user interface~\cite{github}, where parameters for arbitrary hardware configurations can be specified and the token's security following our protocol can be estimated.
	
	While the IBMQ are arguably the most advanced quantum processing units available for public use, this superconducting architectures~\cite{hardware1, hardware2, hardware3} are currently not optimal candidates for such a quantum coin device.
	They operate at low temperatures, making them hardly compact and mobile.
	These results thus pave the way for the application of the protocol with other systems, such as NV centers, even in situations when single shot readout protocols are not feasible.
	Still, certain technological steps need to be overcome in the fabrication of such quantum coins~\cite{token_fabrication}.
	Or as the optimization of control techniques and error mitigation~\cite{QEC_NV} with ensembles, which in the end are most fundamentally concerned with extending the coherence times to application realistic values of more than a few seconds.
	
	The results presented here demonstrate the security of the ensemble-based quantum token protocol under one possible attack scenario, where the hacker performs projective measurements to the bank tokens and attempts to forge a fake coin.	
	Although this attack scenario can be considered a general framework for quantum systems, the demonstrated safety does not imply that the proposed protocol is resilient to all other imaginable attack scenarios.
	This motivates further research into the field, in order that this promising proposal can meet high standards of securities for personal authentication technologies.
	An additional hacking scenario, not considered in Sec.~\ref{sec:attacker}, would be for the attacker to introduce an external entangled qubit to the system and perform a quantum state transfer with the coin tokens, i.e. a SWAP gate~\cite{QIP_NMR}. In this way, quantum no-cloning theorem is not violated, as the state is not cloned but stolen, or in quantum mechanical terms, it is swapped. A concrete example of this is if the attacker uses a scanning tip with an NV~\cite{NV_tip} which can be positioned close to the coin tokens. Such that a hyperfine interaction between the scanning NV and token NVs is established~\cite{NV_entanglement}, generating the means for a SWAP operation. Nonetheless, this technique presents many technical challenges on itself, with a possible fidelity much lower than the bank self-acceptance probability. Additionally, this would herald the cloning once the bank token is read, given by a larger quantum projection noise.
	
	Further questions regarding the security of quantum token devices are outlined in this work.
	Namely, the token ensemble is assumed inseparable, which is trivially achieved in color centers, but may not be the case in other candidate systems.
	Otherwise, if the attacker could make measurements in more than one axis, they would be able to more precisely forge tokens~\cite{OURPRA}.
	This does not invalidate the token protocol, but would merely require more tokens in the coin device.
	Another point which can degrade the protocol's security comes from the fact that our model was developed taking into account pure quantum token states, without considering the intrinsic statistical mixed nature of the ensemble token states.
	Although this approximation is largely valid for the five IBMQ analyzed here, given the high agreement between the theory and experiments, this can potentially not be the case for other ensemble systems as well.
	Where in these cases, the density matrix formalism would need to be used to describe the tokens, with extra variables for the Bloch vector length and uncertainties of the angles.
	
	\begin{acknowledgments}
		
		This work was supported by the German Federal Ministry of Education and Research (BMBF) under the project "\textit{DIamant-basiert QuantenTOKen}" (DIQTOK - n\textsuperscript{o} 16KISQ034).
		
	\end{acknowledgments}

	\section*{Author Contributions}
	
	All authors participated in the discussion, writing and revision of the manuscript. L.T. conceptualized, performed and analyzed the experiments. L.T., B.B., M.X. and K.S. theoretically developed the protocol. M.E.G., K.S. and B.N. conceptualized the quantum token device, acquired funding and supervised the work.
	
	\appendix
	
	
	\section{Mathematical Proofs}\label{sec:A1}
	
	\begin{proof}[Proof of Eq.~\ref{eq:expectation}]\label{proof:expectation_value}
		
		A quantum state with arbitrary angles $\theta$ and $\phi$ is described by
		\begin{equation*}
			\ket{\Psi} = \cos \left(\frac{\theta}{2}\right) \ket{0} + e^{i \phi} \sin \left(\frac{\theta}{2}\right) \ket{1}.
		\end{equation*}
		The expectation value of $\langle \hat{N} \rangle= \bra{\Psi} \hat{N} \ket{\Psi}$ is written as
		\begin{equation*}
			\langle \hat{N} \rangle= 
			\begin{bmatrix}
				\cos \left(\frac{\theta}{2}\right) \\
				e^{i \phi} \sin \left(\frac{\theta}{2}\right) 
			\end{bmatrix} ^T
			\begin{bmatrix}
				N_0 & 0 \\
				0 & N_1 \\
			\end{bmatrix} 
			\begin{bmatrix}
				\cos \left(\frac{\theta}{2}\right) \\
				e^{i \phi} \sin \left(\frac{\theta}{2}\right) 
			\end{bmatrix} .
		\end{equation*}
		Which results in
		\begin{equation*}
			\langle \hat{N} \rangle = N_0 \cos^{2}{\left(\frac{\theta}{2} \right)} + N_1 \sin^{2}{\left(\frac{\theta}{2} \right)}.
		\end{equation*}
	\end{proof}
	
	
	\begin{proof}[Proof of Eq.~\ref{eq:uncertainty}]
		
		Regarding the observable uncertainty $\sigma_N$, we first need to calculate the $\langle \hat{N}^2 \rangle$ term as
		\begin{equation*}
			\langle \hat{N}^2 \rangle =
			\begin{bmatrix}
				\cos \left(\frac{\theta}{2}\right) \\
				e^{i \phi} \sin \left(\frac{\theta}{2}\right) 
			\end{bmatrix} ^T
			\begin{bmatrix}
				N_0^2 & 0 \\
				0 & N_1^2 \\
			\end{bmatrix} 
			\begin{bmatrix}
				\cos \left(\frac{\theta}{2}\right) \\
				e^{i \phi} \sin \left(\frac{\theta}{2}\right) 
			\end{bmatrix}.   
		\end{equation*}
		Which results in
		\begin{equation*}
			\langle \hat{N}^2 \rangle = N_0^2 \cos^{2}{\left(\frac{\theta}{2} \right)} + N_1^2 \sin^{2}{\left(\frac{\theta}{2} \right)} .
		\end{equation*}
		Thus, adding all terms together we get
		\begin{align*}
			\sigma_N^2 & = \sigma_q^2 + \sigma_n^2 + \sigma_{exp}^2 \\
			& = \langle \hat{N}^2 \rangle - \langle \hat{N} \rangle^2 + \langle \hat{N} \rangle + \sigma_{exp}^2 \\
			& = N_0^2 \cos^{2}{\left(\frac{\theta}{2} \right)} + N_1^2 \sin^{2}{\left(\frac{\theta}{2} \right)} \\
			&- \left[N_0 \cos^{2}{\left(\frac{\theta}{2} \right)} + N_1 \sin^{2}{\left(\frac{\theta}{2} \right)}\right]^2 \\
			&+ N_0 \cos^{2}{\left(\frac{\theta}{2} \right)} + N_1 \sin^{2}{\left(\frac{\theta}{2} \right)} + \sigma_{exp} .
		\end{align*}
		Rearranging them results in
		\begin{multline*}
			\sigma_N^2 = N_0 \cos^{2}{\left(\frac{\theta}{2} \right)} (1 + N_0) + N_1 \sin^{2}{\left(\frac{\theta}{2} \right)} (1 + N_1) \\
			- \left[N_0 \cos^{2}{\left(\frac{\theta}{2} \right)} + n_{1} \sin^{2}{\left(\frac{\theta}{2} \right)}\right]^{2} + \sigma_{exp}^2.
		\end{multline*}
		Solutions of $\sigma_N$ for different combinations of $N_0$ and $N_1$ are shown in Fig.~\ref{fig:sigma_analytical}
		
		\begin{figure}[t!]
			\includegraphics[width=\columnwidth]{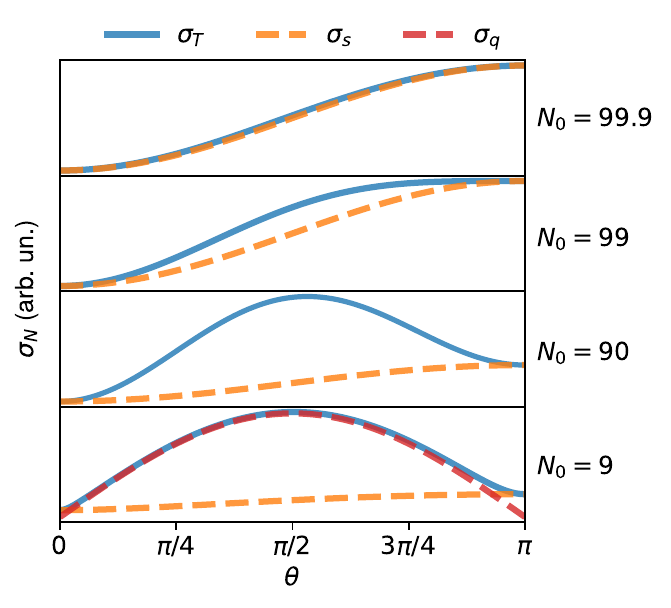}
			\caption{$\sigma_N$ for different values of $N_0$ with fixed $N_1 = 100$ and $\sigma_{exp}=0$. At large $N_0$, the noise is completely dominated by shot noise, which is larger at $\theta=\pi$. On the other hand, as the difference between $N_1$ and $N_0$ increases, the quantum uncertainty becomes the dominating term, being higher at $\theta=\pi/2$.}
			\label{fig:sigma_analytical}
		\end{figure}
		
	\end{proof}
	
	
	\begin{proof}[Proof of Eq.~\ref{eq:na_sol}]
		Now considering the measurement performed by an attacker, we first calculate the final state after a rotation of the bank with $\theta_{b}$ and $\phi_b$, followed by an attacker rotation by $\theta_{a}$ and $\phi_a$. This gives
		\begin{align*}
			\ket{\Psi_f} &= \hat{R}^{-1}(\theta_a, \phi_a )\hat{R}(\theta_b, \phi_b) \ket{0} \\
			& =
			\begin{bmatrix}
				\cos \left(\frac{\theta_a}{2}\right) & i \sin \left(\frac{\theta_a}{2}\right) e^{-i \phi_a} \\
				i \sin \left(\frac{\theta_a}{2}\right) e^{i \phi_a}  & \cos \left(\frac{\theta_a}{2}\right) \\
			\end{bmatrix} \\
			& \times
			\begin{bmatrix}
				\cos \left(\frac{\theta_b}{2}\right) &  -i \sin \left(\frac{\theta_b}{2}\right) e^{-i \phi_b} \\
				-i\sin \left(\frac{\theta_b}{2}\right) e^{i \phi_b}  & \cos \left(\frac{\theta_b}{2}\right) \\
			\end{bmatrix} 
			\begin{bmatrix}
				1 \\
				0 \\
			\end{bmatrix} \\
			& =
			\begin{bmatrix}
				\cos \left(\frac{\theta_a}{2}\right) & i \sin \left(\frac{\theta_a}{2}\right) e^{-i \phi_a} \\
				i \sin \left(\frac{\theta_a}{2}\right) e^{i \phi_a}  & \cos \left(\frac{\theta_a}{2}\right) \\
			\end{bmatrix}
			\begin{bmatrix}
				\cos \left(\frac{\theta_b}{2}\right) \\
				-i \sin \left(\frac{\theta_b}{2}\right) e^{i \phi_b} \\
			\end{bmatrix} \\ 
			& =
			\begin{bmatrix}
				\cos \left(\frac{\theta_a}{2}\right)\cos \left(\frac{\theta_b}{2}\right) + \sin \left(\frac{\theta_a}{2}\right)\sin \left(\frac{\theta_b}{2}\right) e^{i (\phi_b-\phi_a)}  \\
				i\sin \left(\frac{\theta_a}{2}\right)\cos \left(\frac{\theta_b}{2}\right) e^{i \phi_a} - i \sin \left(\frac{\theta_b}{2}\right)\cos \left(\frac{\theta_a}{2}\right) e^{i \phi_b}\\
			\end{bmatrix} ,
		\end{align*}
		where we use the rotation operator as defined in~\cite{QIP_NMR}. The expectation value of $\hat{N}$ for this state is
		\begin{align*}
			N_0 \left| \cos \left(\frac{\theta_a}{2}\right)\cos \left(\frac{\theta_b}{2}\right) + \sin \left(\frac{\theta_a}{2}\right)\sin \left(\frac{\theta_b}{2}\right) e^{i (\phi_b-\phi_a)} \right|^2 \\
			+ N_1 \left|\sin \left(\frac{\theta_a}{2}\right)\cos \left(\frac{\theta_b}{2}\right) e^{i \phi_a} - \sin \left(\frac{\theta_b}{2}\right)\cos \left(\frac{\theta_a}{2}\right) e^{i \phi_b} \right|^2 .
		\end{align*}
		Using basic trigonometric relations for $\cos (2x)$ and $\sin (2x)$, the term with $N_0$ gives
		\begin{align*}
			\cos^2 \left(\frac{\theta_a}{2}\right)\cos^2 \left(\frac{\theta_b}{2}\right) + \sin^2 \left(\frac{\theta_a}{2}\right)\sin^2 \left(\frac{\theta_b}{2}\right) \\
			+ \cos \left(\frac{\theta_a}{2}\right)\cos \left(\frac{\theta_b}{2}\right) \sin \left(\frac{\theta_a}{2}\right)\sin \left(\frac{\theta_b}{2}\right) 2 \cos (\phi_b-\phi_a) \\
			= \frac{1}{2} \left[ 1 + \cos \theta_a \cos \theta_b + \sin \theta_a \sin \theta_{b} \cos (\phi_b-\phi_a) \right] ,
		\end{align*}
		while the term with $N_1$ yields
		\begin{align*}
			\sin^2 \left(\frac{\theta_a}{2}\right)\cos^2 \left(\frac{\theta_b}{2}\right) + \sin^2 \left(\frac{\theta_b}{2}\right)\cos^2 \left(\frac{\theta_a}{2}\right) \\
			-\sin \left(\frac{\theta_a}{2}\right)\cos \left(\frac{\theta_b}{2}\right) \sin \left(\frac{\theta_b}{2}\right) \cos \left(\frac{\theta_a}{2}\right) 2 \cos (\phi_b-\phi_a) \\
			= \frac{1}{2} \left[ 1 - \cos \theta_a \cos \theta_b - \sin \theta_a \sin \theta_{b} \cos (\phi_b-\phi_a) \right] .
		\end{align*}
		Adding both terms together we get
		\begin{equation*}
			\frac{N_0 + N_1}{2} + \frac{N_0-N_1}{2}[\cos \theta_a \cos \theta_b  + \sin \theta_a \sin \theta_{b} \cos (\phi_b-\phi_a)  ] \\
		\end{equation*}
		Finally, the fraction of qubits measured in the $\ket{0}$ state by the attacker $n_a$ will be
		\begin{equation*}
			1- n_a = \frac{\bra{\Psi_f} N \ket{\Psi_f}}{N_0 + N_1} ,
		\end{equation*}
		which leads to
		\begin{multline*}
			2 n_a = 1 + \frac{N_1-N_0}{N_0+N_1} [ \cos{\theta_a}\cos{\theta_b} \\ + \sin{\theta_a}\sin{\theta_b}\cos{(\phi_b - \phi_a)} ] .
		\end{multline*}
	\end{proof}
	
	
	\begin{proof}[Proof of $\bar{n}_a = 1/2$]
		To get the average value of $\bar{n}_a$, we integrate Eq.~\ref{eq:na_sol} over the whole Bloch sphere
		\begin{multline*}
			2 \bar{n}_a = \int_{0}^{2 \pi} d\phi_b \int_{0}^{\pi} \frac{\sin \theta_{b}}{2} d\theta_b \\ \times \left\{ 1 + c [ \cos{\theta_a}\cos{\theta_b} + \sin{\theta_a}\sin{\theta_b}\cos{(\phi_b - \phi_a)} ] \right\} ,
		\end{multline*}
		The first term simply gives 1, while the two other result in
		\begin{multline*}
			2 \bar{n}_a = 1  + \frac{c}{2}\cos \theta_a \int_{0}^{\pi} \sin \theta_{b}\cos \theta_{b} d\theta_b \\ + \frac{c}{2}\sin \theta_a \int_{0}^{\pi} \sin^2 \theta_{b} d\theta_b \int_{0}^{2 \pi} \cos{(\phi_b - \phi_a)} d\phi_b .
		\end{multline*}
		It is straight forward to see that both integrals in the second and third term result in 0. Thus, the final average value of $n_a$ is simply
		\begin{equation*}
			\bar{n}_a = \frac{1}{2} .
		\end{equation*}
		
	\end{proof}
	
	
	\begin{proof}[Interval of valid solutions for $\theta_f$]
		
		\begin{figure}[t!]
			\includegraphics[width=\columnwidth]{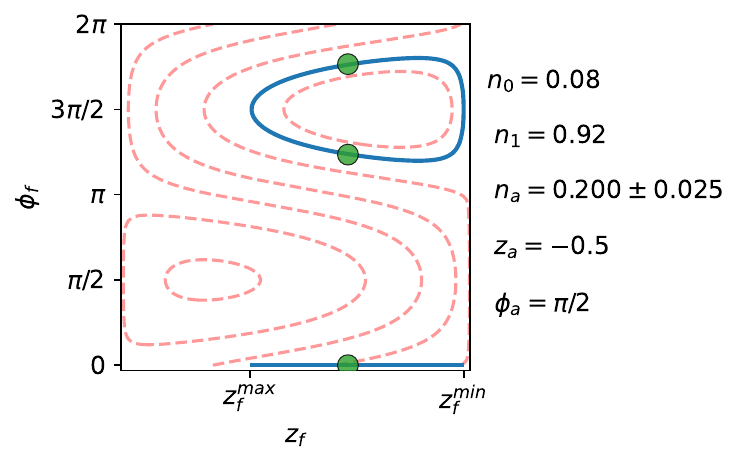}
			\caption{Solution for forger angles $\theta_f$ and $\phi_f$ with fixed $n_a$ inside the interval $0.200\pm0.025$, depicting how the attacker can forge a fake token based on their measurement of the bank token. First, the attacker calculates the interval solution for $z_f$ and then randomly takes a value inside the interval. Last, they solve Eq.~\ref{eq:na_sol} for $\phi_f$ and choose one of the two possible solutions.}
			\label{fig:zf_interval}
		\end{figure}
		
		In order for the parametric relation from Eq.~\ref{eq:na_sol} to have real solution for $\phi_f$ we need that the argument inside the inverse cosine function is in the interval $[-1,1]$. Defining $\alpha \equiv (2n_a-1)/c$, this results in
		\begin{align*}
			\left( \alpha - \cos \theta_a \cos \theta_f \right)^2 &< \left( \sin \theta_a \sin \theta_f \right)^2 \\
			\alpha^2 - 2 \alpha \cos \theta_a \cos \theta_f + \cos^2 \theta_a \cos^2 \theta_f &-\sin^2 \theta_a \sin^2 \theta_f < 0 .
		\end{align*}
		Again, using the $\cos(2x)$ relation we have
		\begin{equation*}
			\alpha^2 - 2 \alpha \cos \theta_a \cos \theta_f + \frac{\cos{2\theta_{a}}}{2} + \frac{\cos^2\theta_{f} - \sin^2\theta_{f} }{2} <0 .
		\end{equation*}
		Now we can eliminate the dependency on $-\sin^2 \theta_f$ substituting it by $\cos^2 \theta_f - 1$ and get to a quadratic equation for $z_{f} = \cos \theta_{f}$ as
		\begin{equation*}
			z_f^2 + z_f \left( -2\alpha \cos \theta_{a} \right) + \left ( \alpha^2 + \frac{\cos{2\theta_{a}}}{2} - \frac{1}{2} \right) >0 .
		\end{equation*}
		Using a simple quadratic formula solution this results in interval of solution for $z_{f}$ of
		\begin{multline} \label{eq:zf_interval}
			\max \left\{ \alpha \cos \theta_a  - \sqrt{\Delta}, -1 \right\} \leqslant z_{f} \\ \leqslant \min \left\{ \alpha \cos \theta_a  + \sqrt{\Delta}, 1 \right\} ,
		\end{multline}
		with
		\begin{equation*}
			\Delta = \alpha^2 \cos^2 \theta_a - \alpha^2 - \frac{\cos{2\theta_{a}}}{2} + \frac{1}{2} .
		\end{equation*}
		The solution interval for $z_f$ is depicted in Fig.~\ref{fig:zf_interval}
		
	\end{proof}
	
	
	\begin{figure*}[t!]
		\includegraphics[width=\textwidth]{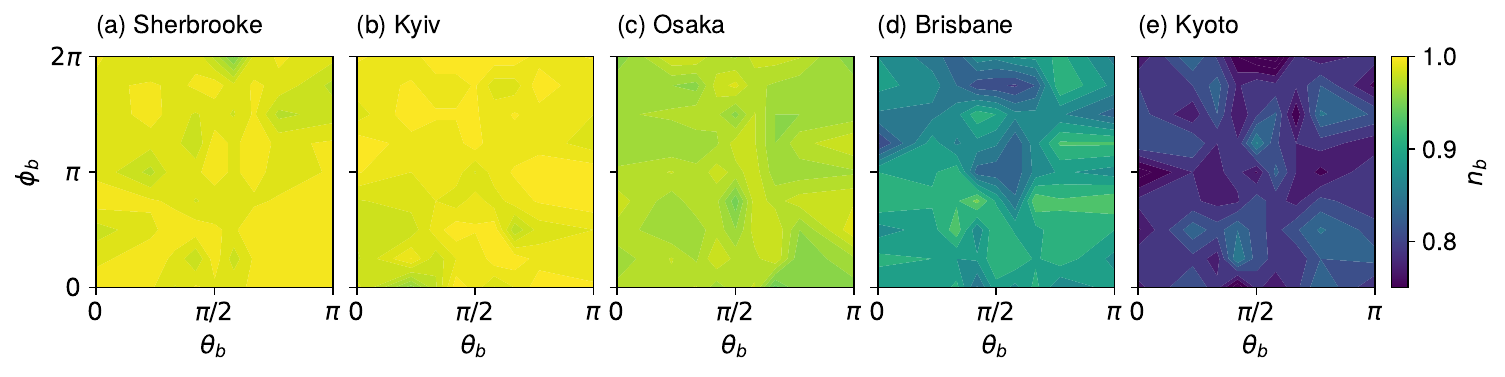}
		\caption{Angle dependence of the self-acceptance fraction $n_b$ for IBM Quantum Platforms \textbf{(a)} Sherbrooke, \textbf{(b)} Kyiv, \textbf{(c)} Osaka, \textbf{(d)} Brisbane and \textbf{(e)} Kyoto. In hardware with unoptimized control and short coherence times, smaller $n_b$ would be expected for larger angles due to prolonged pulses. Which is not the case here for the IBMQ.}
		\label{fig:nb_angles}
	\end{figure*}
	
	\section{Experimental Methods}\label{sec:A2}
	
	To fully benchmark the protocol and validate our model, measurements were performed in five distinct IBMQ of the Eagle family: Sherbrooke, Kyiv, Brisbane, Kyoto (retired) and Osaka (retired). All of them have similar superconducting architecture~\cite{hardware1, hardware2, hardware3}, with 127 physical qubits and the same native operations. Still, their exact specifications are not publicly available. Comparatively, Sherbrooke presented the longest mean coherence time $T_2$, while Brisbane had the smallest gate error and Kyiv the smallest readout error, based on IBM calibration reports. Kyoto, on the other hand, had the worse parameters of the five. Small changes in these performance parameters were observed during the experiments conducted from April 2024 to December 2024, but without much implication to the results here presented. Due to high memory costs for waveform generation and increased gate and readout errors, an ergodic approximation was used for the qubit ensemble, where a single qubit was averaged over time instead of using multiple qubits at once. The use of such an ergodic approximation is further validated by the assumption of the inseparability of the ensemble.
	
	Experiments were run in the hardware through Qiskit software development tool~\cite{qiskit} in Python, version 1.1.1 and older. All Qiskit codes used for measurements and data analysis are open-source, provided at the author's GitHub repository~\cite{github}. The high-level representation of the quantum circuits were defined with the Quantum Circuit class. Which were then transpiled to the hardware's native gates without gate approximation, using the qubit with longest $T_2$ and shortest gate operation, while considering the specific timing constrains of each quantum hardware. Finally, the protocols were run with Qiskit Runtime, while the corresponding observable $\hat{N}$ expectation values are taken from the job counts in the $\ket{0}$ and $\ket{1}$ states. 
	
	\begin{table*}[t!]
		\begin{tabular}{|c c|}
			\hline
			Variable & Physical Meaning \\ \hline
			$\theta$ & polar angle on the Bloch sphere \\
			$\phi$ & azimuthal angle on the Bloch sphere \\
			$z$ & $\cos \theta$ \\
			$\theta_{b}, \phi_b$ & angles which the bank prepares and measures the token \\ 
			$\theta_{a}, \phi_a$ & angles used by the attacker to measure the bank token \\
			$\theta_{f}, \phi_f$ & angles forged by the attacker \\
			
			$\hat{N}$ & system's observable (Example: photon count) \\
			$N_0$ & $\hat{N}$ eigenvalue for $\ket{0}$ (Example: dark counts) \\
			$N_1$ & $\hat{N}$ eigenvalue for $\ket{1}$ (Example: bright counts)  \\
			$c$ & normalized contrast between $N_1$ and $N_0$ \\
			$\sigma_N$ & total uncertainty of $\hat{N}$ \\
			$\sigma_q$ & Heisenberg uncertainty \\
			$\sigma_s$ & shot noise \\
			$\sigma_{exp}$ & experimental noise \\
			
			$\hat{R}(\theta, \phi)$ & rotation operator with angles $\theta$ and $\phi$ \\
			
			$n$ & fraction of qubits measured in the $\ket{0}$ state \\
			$n_b$ & $n$ if the bank prepares and measures without an attacker \\
			$n_a$ & $n$ measured by the attacker with angles $\theta_{a}, \phi_a$ for a bank token \\
			$n_f$ & $n$ measured by the bank for a forged token \\
			$n_r$ & $n$ measured by the bank for randomly forged tokens \\
			$n_T$ & minimum $n$ threshold for the token to be accepted by the bank \\
			
			$M$ & number of tokens in the device \\
			$p_b$ & self-acceptance probability of the bank tokens \\
			$p_f$ & acceptance probability of forged tokens \\
			
			$\alpha$ & $(2n_a-1)/c$ \\
			\hline
		\end{tabular}
		\caption{Glossary of the main variables in the text.}
		\label{tab:variables}
	\end{table*}
	
	\begin{figure*}[t!]
		\centering
		\includegraphics[width=\textwidth]{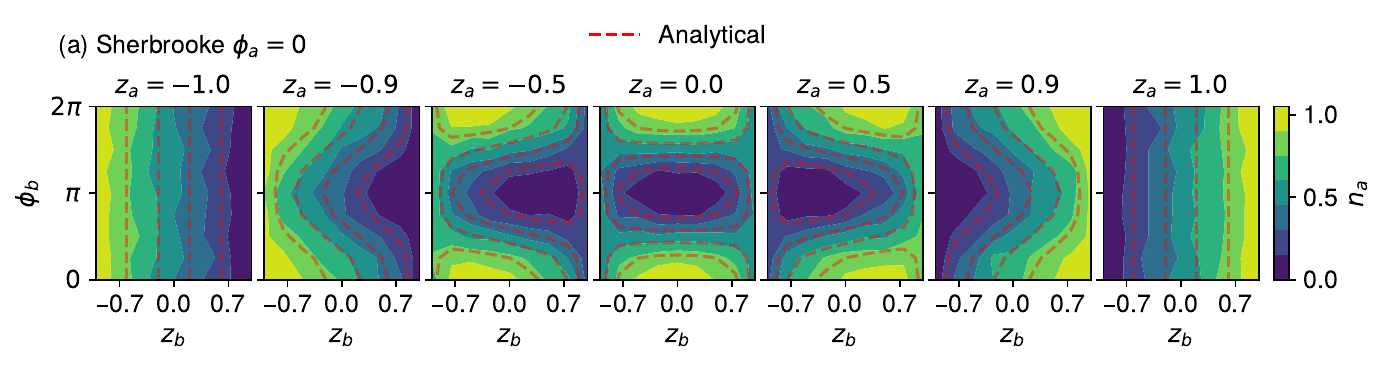}\vspace{-.2cm}
		\includegraphics[width=\textwidth]{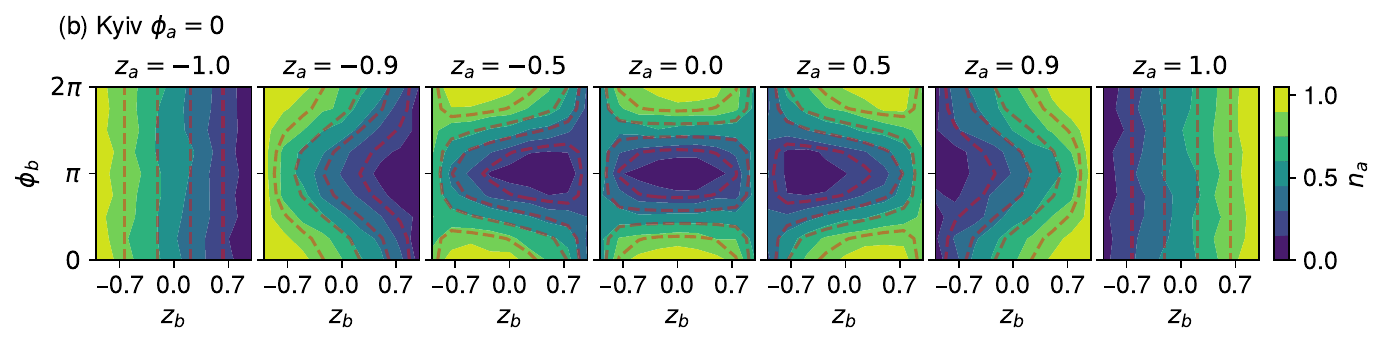}\vspace{-.2cm}
		\includegraphics[width=\textwidth]{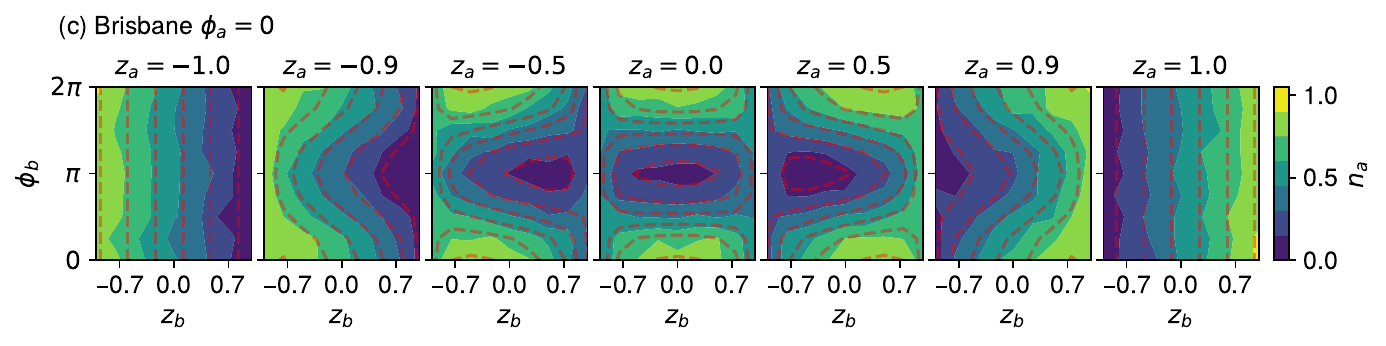}\vspace{-.2cm}
		\includegraphics[width=\textwidth]{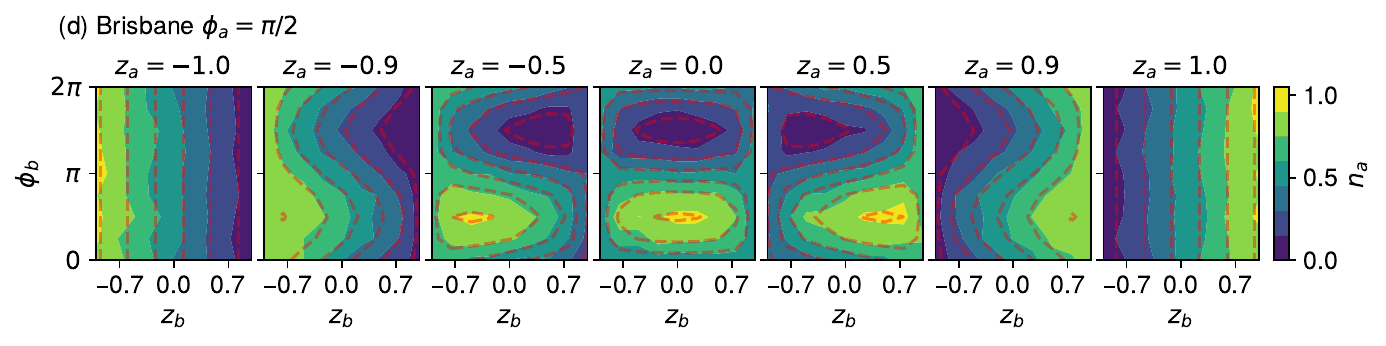}\vspace{-.2cm}
		\includegraphics[width=\textwidth]{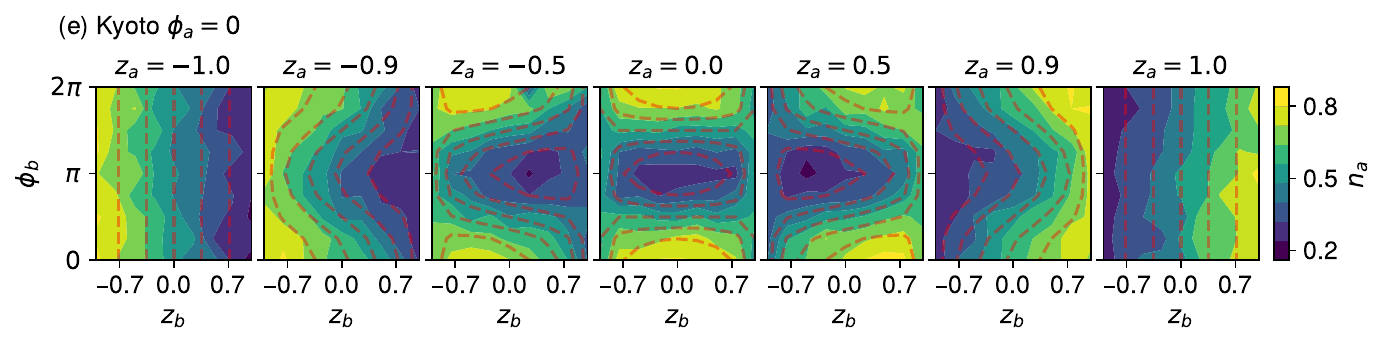}
		\caption{Angle dependence of $n_a$ measured with $\phi_a=0$ in IBM Quantum Platforms \textbf{(a)} Sherbrooke, \textbf{(b)} Kyiv, \textbf{(c)} Brisbane and \textbf{(e)} Kyoto, in addition to \textbf{(d)} Brisbane measured with $\phi_a=\pi/2$. All of them are compared with the analytical solution from Eq.~\ref{eq:na_sol} of (red dashed lines). All hardware show strong agreement with the analytical formula, except Kyoto which has more noise due to its increased experimental uncertainty, as well as a smaller contrast (Tab.~\ref{tab:fit_params}). Furthermore, the attacker angle of $\phi_a=\pi/2$ in Brisbane simply shifts the angle dependence of $n_a$ compared to $\phi_a=0$.}
		\label{fig:na_SM}
	\end{figure*}
	
	\clearpage

\end{document}